\documentclass{ws-ijmpb}
\begin{document}

\markboth{Boyer and Romeu}
{Nonlinear and geometrical theories of grain boundaries}

%
\catchline{}{}{}{}{}
%

\title{MODELING GRAIN BOUNDARIES IN SOLIDS USING A COMBINED
NONLINEAR AND GEOMETRICAL METHOD}

\author{DENIS BOYER}

\address{Instituto de F\'\i sica, Universidad
Nacional Aut\'onoma de M\'exico,\\ Apartado Postal 20-364,
01000 M\'exico D.F., M\'exico\\
boyer@fisica.unam.mx}

\author{DAVID ROMEU}

\address{Instituto de F\'\i sica, Universidad
Nacional Aut\'onoma de M\'exico,\\ Apartado Postal 20-364,
01000 M\'exico D.F., M\'exico\\
romeu@fisica.unam.mx}

\maketitle

\begin{history}
\received{Day Month Year}
\revised{Day Month Year}
\end{history}

\begin{abstract}
The complex arrangements of atoms near grain
boundaries are difficult to understand theoretically.
We propose a phenomenological (Ginzburg-Landau-like) description of
crystalline phases based on symmetries and fairly general
stability arguments.
This method allows a very detailed description of defects at
the lattice scale with virtually no tunning parameters, unlike
usual phase-field methods. The model equations
are directly inspired from those used in a very different physical
context, namely, the formation of periodic patterns
in systems out-of-equilibrium ({\it e.g.} Rayleigh-B\'enard convection,
Turing patterns).
We apply the formalism to the study of symmetric tilt boundaries. 
Our results are in
quantitative agreement with those predicted by a recent crystallographic
theory of grain boundaries based on a geometrical quasicrystal-like
construction.
These results suggest that frustration and competition effects near
defects in crystalline arrangements have some universal
features, of interest in solids or other periodic phases.
\end{abstract}

\keywords{Grain boundaries; Pattern formation; Geometrical frustration; 
Quasicrystals; Faceting.}

\section{Introduction}
The structure and dynamics of grain boundaries in materials
are still little understood.\cite{1,2}
The possible relationships between the mesoscopic features
of grain boundaries and their inner
microscopic structure at the atomic scale remain unclear.
The energies of some interfaces can be accurately described with
continuous theories and topological arguments. But the study of other
quantities (like mobilities), as well as short scale phenomena such
as faceting, may require a deeper understanding of the microscopic
structure, which directly
reflects the discrete and anisotropic character of the lattice.
Near a boundary, the crystal structure usually has no simple periodic behavior
and is therefore difficult to characterize.
Any quantitative and sufficiently simple theory allowing a systematic
characterization of the fine structure of grain boundaries at zero
temperature could open the way toward a general description of these objects.
Following such an approach requires to go beyond usual continuous
media approximations.

In this paper, we discuss a way of elucidating grain boundary
microscopic structures within a theoretical framework that
explicitly takes into account spatial variations at the atomic length
scale. The models considered here have a very small number of
parameters. They derive from very general energy and symmetry
arguments. Although the nonlinear approach proposed below is
fairly uncommon for crystals, it is largely used in a (apparently)
completely different field of Physics, namely, in pattern
formation problems. Many physical systems driven out of
equilibrium spontaneously form periodic structures, some of which
are characterized by a crystal-like order.\cite{3} Some
well-known examples are the cellular structures formed in
Rayleigh-B\'enard convection of heated fluid layers, 
or Turing patterns in reaction-diffusion processes. Pattern formation is 
ubiquitous in
nature and has universal features. Some of these features may be
of relevance for solid crystals as well. Striking analogies
between pattern forming systems and dissipative processes taking
place in solid crystals have been recently pointed 
out.\cite{4,5} In Section \ref{sec2}, we briefly
justify why some simple nonlinear models of pattern formation
may be useful to study defects in
crystals. We show in Sec. \ref{sec3} that these models are very
efficient to capture elementary competition and geometrical
exclusion effects that probably take place in genuine grain
boundaries. In particular, they are able to predict interface
faceting, a feature still little understood from continuous
theories.\cite{6} The results obtained are in very
good agreement with those given independently by a theory of grain
boundaries developed recently, which is based on geometrical
arguments and inspired from quasicrystals theory.\cite{7}

\section{Basics of pattern formation}
\label{sec2}

We briefly recall some concepts of pattern formation, that can be
found in well documented reviews or textbooks.\cite{3,8,9}
Consider a spatially extended system described by a
phenomenological, dimensionless local order parameter $\psi(\vec{x},t)$
depending {\em a priori} on space and time variables.
In convection problems, for
instance, $\psi(\vec{x},t)$ is related to the vertical velocity at the 
mid-plane of the convective cell. A uniform phase is characterized by the
trivial solution $\psi=$cst($=0$). When a non-trivial stationary periodic
pattern sets up in the system, {\it e.g.} above the onset of convection,
$\psi$ can be written in a first approximation as a sum of plane waves:
\begin{equation}\label{sum}
\psi(\vec{r})=\sum_{n=1}^{N}A_n\cos(\vec{k}_n\cdot\vec{x}),
\end{equation}
where $N$ is the number of wavevectors (of same modulus
$|\vec{k}_n|=k_0=2\pi/\lambda_0$) characterizing the base pattern
(in two spatial dimensions, $N=3$ for hexagonal symmetry and 
$N=2$ for square symmetry).
A perfect crystal is observed when the amplitudes
$A_n$ are all equal to a constant $A_0$. The
Figures \ref{low} and \ref{high}
further show examples of the field $\psi$ in gray scale in 2D 
(dark regions have $\psi(\vec{x})\ge 0$, bright ones $\psi\le0$).
We now look for a simple phenomenological isotropic partial differential
equation satisfied by $\psi$. Keeping in mind the analogy with crystals,
in addition to the periodicity $\lambda_0$, one wishes an equation that
{\em imposes a symmetry} on the stationary solution (for instance hexagonal,
$N=3$). A linear wave equation like
$(\Delta+k_0^2)\psi=0$ is not acceptable, as a superposition
of any $N\ge 1$ waves would be solution.
Instead, we need to resort to a nonlinear equation, introducing a
coupling between the different amplitudes $A_n$ in Eq.(\ref{sum}).
We consider here the well-known Swift-Hohenberg model
(first introduced in the context of Rayleigh-B\'enard convection),
which also involve the time variable $t$:\cite{3}
\begin{equation}
\label{sh}
\frac{\partial \psi}{\partial t}=\epsilon \psi-\frac{1}{k_0^4}
(k_0^2+\Delta)^2\psi+g_2\psi^2-\psi^3,
\end{equation}
where $\epsilon\ll 1$ is a dimensionless control parameter, the
reduced Rayleigh number in convection problems.
If $\epsilon\le0$ in (\ref{sh}), $\psi=0$ is the only stationary
solution ($\partial \psi/\partial t=0$) that is stable. The uniform solution
$\psi = 0$ becomes unstable for $\epsilon > 0$. In that case, new solutions
with broken symmetries appear, in the form of a periodic pattern characterized
by a layer spacing $\lambda_{0} = 2\pi/k_{0}$. The order parameter
$\psi$ can be approximated by a sum of the form (\ref{sum})
provided that $\epsilon\ll1$.
Nonlinear terms are responsible for mode combinations: given an
arbitrary set of wavevectors $\{\vec{k}_n\}$, one can derive equations
for their amplitudes $A_n$, these amplitudes being now coupled to
each other. The $A_n$'s happen to have stable non-zero solutions only in a
very small number of cases, that generally
correspond to symmetric, crystal-like arrangements of the $\{\vec{k}_n\}$.
For instance, equation (\ref{sh}) with $g_2=0$ leads to stripes formation
($N=1$). On the other hand, if $g_2\neq 0$ and $0<\epsilon< 4g_2^2/3$ 
only hexagonal solutions are stable.\cite{9}

Let us now point out one of the crucial properties of this type of model:
the evolution equation (\ref{sh}) has a {\em potential structure}.
One can associate an \lq\lq energy", or Liapunov functional $F$,
to the system, such that Eq.(\ref{sh}) reads
$\partial \psi/\partial t=-\delta F/\delta\psi$
(the right-hand-side representing a functional
derivative). $F$ is given in that case by
\begin{equation}\label{F}
F=\int d\vec{x}\left[-\frac{\epsilon}{2}\psi^2
+\frac{\psi}{2k_0^4}(k_0^2+\Delta)^2\psi-g_2\frac{\psi^3}{3}+\frac{\psi^4}{4}
\right].
\end{equation}
It is easy to show from Eq.(\ref{sh}) that
$dF/dt\le 0$: the quantity $F$
always decreases with time. The dynamics is dissipative and converges
toward stationary stable solutions that correspond to
{\em local minima} of $F$ in the configuration space.
The absolute minimum of $F$ corresponds a perfectly regular
pattern of the form (\ref{sum}), with wavenumber $k_0$
(a \lq\lq perfect crystal"). Consider now an immobile grain boundary
separating two differently oriented domains that are otherwise in the ground
state: near the boundary, it is impossible to preserve the regular crystal
structure imposed in the bulk. As a consequence, such a configuration must
have a higher $F$.

One can now see the pattern formation process given by (\ref{sh}) from
an alternate view-point, very useful for applications to materials:
namely, as an {\em energy minimization} problem within some constraints.
Stationary grain boundaries can be regarded as optimal solutions
of a frustration problem of two competing lattices of different orientations.
The free-energy functional imposes a fixed {\em periodicity} on the field
$\psi$ in the bulk, minimizing the term $\psi(k_0^2+\Delta)^2\psi$ in
(\ref{F}), as well as a given bulk {\em symmetry}, which is fixed by mode
combination through the higher order terms. 
The equations for the amplitudes $A_n$ show that
the grain boundary is actually an interference
pattern between the two crystals, with amplitudes not constant
in space and not equal to each other.

Different choices of nonlinear terms can lead to different crystal
symmetries in the bulk. Adding other cubic or higher order terms in the
functional
(\ref{F}) generates new nonlinear terms in Eq.(\ref{sh}), modifying
the relative coupling constants between the different amplitudes $A_n$
in the solution (\ref{sum}). As a consequence, a crystal formerly
unstable may become stable. For instance, a square crystal lattice
($N=2$) can be obtained by choosing:\cite{10}
$F=\int d\vec{x}[-(\epsilon/2)\psi^2
+1/(2k_0^4)\psi(k_0^2+\Delta)^2\psi-g_2\psi^3/3
+\beta\psi^4/4+(\gamma/4)\psi^2\Delta^2(\psi^2)]$, with $\beta<0$
and $\epsilon>16g_2^2(3\beta+13\gamma)/(3\beta+4\gamma)^2$.
This free energy gives coefficients that depend on the angle
between the wavevectors in the amplitude equations, such that they stabilize
solutions with perpendicular wave numbers.

The energy (\ref{F}) involves a Taylor series expansion of
an order parameter, as usual in Ginzburg-Landau descriptions of phase
transitions. This simple form suggests that geometrical
frustration in crystalline phases with isotropic interactions is likely
to have some universal features. Actually, numerical solutions of
equation (\ref{sh}), or slight variants of it, are able to reproduce
known features of crystals. The behavior of the energy $F$ of grain boundaries
as a function of misorientation angle follows a
Read-Shockley law.\cite{4} In addition, short-range pinning
forces act on defects, a phenomenon analogous to the Peierls stress
in crystals.\cite{5}

In the following
Section, we emphasize the close relationships that exist between
this pattern formation approach of grain boundaries and some space filling
problems under constraints, that have been solved in Ref.~\refcite{7} by
the use of quasi-crystal methods.
We briefly recall the main aspects of this second theory below.

\section{Structures of grain boundaries}
\label{sec3}

\subsection{A geometrical construction}
\label{david}
The main distinctive feature of this theory is that it considers
interfaces
and quasicrystals as a region in space where two or more interpenetrating
crystal lattices compete for space. The final atomic positions are then
decided by a modified version of the strip-projection proposed by Katz and
Duneau\cite{11,12} to study quasicrystals. Perhaps
the most important attribute of this approach is that it allows interfaces and
quasicrystals to be described by the same set of equations, thus rendering
them formally equivalent.

Given a completely arbitrary set of lattices and relative orientations the
formalism produces ideal (minimum local strain) structures which are expected
to play for grain boundaries the same role than the perfect crystal concept
plays for crystals, i.e., they define defect free, reference structures
analogous to Bravais lattices. Since strain minimization is a physical
consideration, the formalism is endowed with a physical basis in spite of its
geometrical formulation.

Full details of this method can be found in Ref.~\refcite{7}.
Briefly, the crystal lattices are embedded in a higher dimensional space
where they define an hyper lattice whose symmetry depends on that of the
lattices and their relative orientation. This hyper lattice is
then projected back to physical space to produce the actual interfacial
structure or quasicrystal.

\begin{figure}[bt]
\centerline{\psfig{file=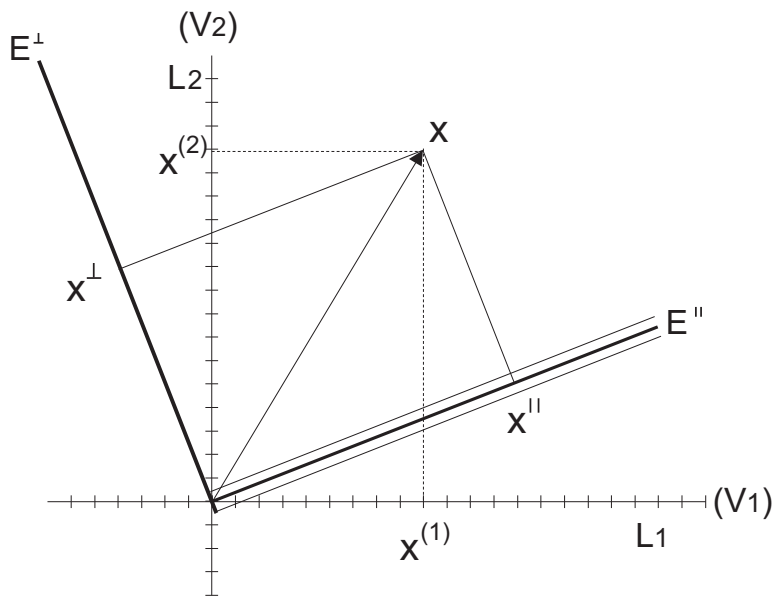,width=4.in}}
\vspace*{8pt}
\caption{\label{hyper} Schematic representation of 6D space showing the
orthogonal 3d subspaces
$V_{1},V_{2},E^{\parallel},E^{\perp}$ as 1D lines. The crystal lattices $L_{1}$
and $L_{2}$ are respectively contained in $V_{1},V_{2}$. The interface is
defined as the set of points $x^{\parallel}$ that are projections of the
hyperlattice points located inside a small region around $E^{\parallel}$.
This region is called the strip and is delimited by the two thin
lines (see text).}
\end{figure}

The higher dimensional $6D$ space is schematically represented in
Figure \ref{hyper}. It is divided in two sets of orthogonal $3D$
subspaces: $(V_{1},V_{2})$ and $(E^{\parallel},E^{\perp})$.
$V_{1}$ and $V_{2}$ contain the lattices $L_{1}$
and $L_{2}$ of the two crystals, and $E^{\parallel},\ E^{\perp}$ 
contain respectively, the actual interfacial structure and the strain 
field across the interface. $E^{\perp}$ is actually the Bollmann's 
displacement space or b-space.\cite{1}

Only those hyperpoints contained within a small region around
$E^{\parallel}$ (see Figure \ref{hyper}) called the strip\cite{11,12}
are projected. The size of the strip is
chosen to include pairs of atoms that occupy incompatible
positions, these atoms are then replaced in $E^{\parallel}$ by a
single atom at their average position. Given two
interpenetrating lattices in physical space, the method constructs
an ideal, best fit, minimum strain lattice as the set of points of
$E^{\parallel}$. By adopting the average position, atoms at the
interface act as a strain buffer between the crystals on each side
of the interface. The component in $E^{\perp}$ is a measure of the
local strain or frustration between two nearly coincident
(overlapping) atoms from each lattice.

This method allows the construction of a complete bicrystal with
the parent crystals separated by a unique, fully 3D structure
called the interfacial layer (IL), arising from the projection of
hyperpoints within the strip. The role of the IL, whose structure
is different from that of either crystal is to resolve the
positional incompatibilities that occur at the interface. Since
grain boundaries are usually 2D, a suitable IL plane (or surface)
must be selected in order to model any particular boundary plane,
of given relative orientation to the parent
crystals. A bicrystal can then be built by filling the space on
either side of the boundary with crystal sites projected from
$(V_{1})$ and $(V_{2})$. Note this approach is unique in that the
grain boundary always contains atomic (IL) sites, which makes it
susceptible of crystallographic analysis.

We recall now the definition of the O-lattice, a concept that will be useful
in the next Section.
Given two crystal lattices, $L_{1}$, $L_{2}$ and a transformation $\mathbf{A}%
$, such that $L_{2}={\mathbf A}L_{1},$ in the absence of translations, the
O-lattice is defined as the set of points $x^{(o)}$ that satisfy
\begin{equation}
\left(  {\mathbf I}-{\mathbf A}^{-1}\right)  ^{-1}l^{(1)}=x^{(o)}%
\label{Bollmann equation}%
\end{equation}
where $\mathbf{I}$ is the identity matrix, and $l^{(1)}\in$
$L_{1}.$ Geometrically, the O-lattice corresponds to points in
space (not necessarily lattice points) of minimum strain i.e.,
minimum misfit between $L_{1}$ and $L_{2}$. Another interesting
property of the O-lattice is that if $l^{(i)\ast}$ and
$\left({\mathbf A} l^{(i)}\right)^{\ast}$ with $i=1,2,3$ are the
base vectors of the reciprocal lattices of
$L_{1}$ and $L_{2}$ respectively, then %

\begin{equation}\label{OLperio}
\left(  l^{(i)\ast}-\left({\mathbf A} l^{(i)}\right)^{\ast}
\right) \cdot x^{(o)}=n
\end{equation}
with integer $n.$ This equation establishes a relationship between
the O-lattice and the Moir\'e
pattern formed by the superposition of $L_{1}$ and $L_{2}$.

\subsection{Results}

In this section we study symmetric tilt grain boundaries in two-dimensional
($2D$) crystals.
The initial condition $\psi(\vec{x},t=0)$ in Eq.(\ref{sh}), or its analogue
for square patterns, is made of two crystals rotated by $\theta_{mis}$,
the interface being
directed along the vertical axis (see Ref.~\refcite{5} for more details).
We introduce an angle $\theta$ that either corresponds to the misorientation
angle $\theta_{mis}$ or to $90^\circ-\theta_{mis}$ (for square patterns),
depending on the orientation of the symmetric boundary plane. Note that
$0\le\theta_{mis}\le 45^\circ$ and $0\le\theta\le 90^\circ$.
For each rotation $\theta_{mis}$, there are two possible boundary
orientations (with different structure) that lead to a symmetric interface.
The angle $\theta$ represents the angle formed between two atomic planes that
are symmetric with respect to the boundary.

We analyze the stationary (minimum energy) fields obtained
at large times by solving Eq.(\ref{sh}) numerically, and
compare their structures with those given by
the method described in Sec. \ref{david}.

\begin{figure}[bt]
\centerline{\psfig{file=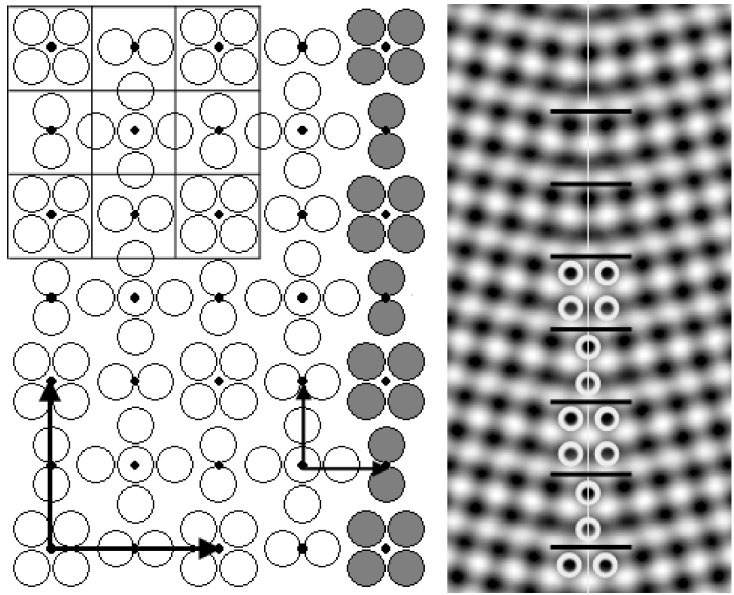,width=4.in}}
\vspace*{8pt}
\caption{\label{low} Singular tilt boundary with $\theta=28.072^{\circ}$
($\Sigma17$). Right: $\psi$
in gray scale given by the nonlinear model.
Left: $E^{\parallel}$ from the construction of Sec. \ref{david}. See text
for legend.}
\end{figure}

Figure \ref{low} displays (on the right) a boundary obtained by
solving the Swift-Hohenberg equation for square patterns, with an
angle $\theta=2\ {\rm tan}^{-1}(1/n)$, where $n$ is an integer (4
in this case). Such interfaces are singular\cite{7} and
characterized by a simple repeating pattern, as emphasized by the
bright circles.

In the geometrical method, any grain boundary structure is a 2D
manifold (a plane, usually) included in the 3D IL. The most likely
manifolds are {\it a priori} the ones that run along directions that maximize 
the density of IL sites. Such dense planes must clearly intersect
perpendicular IL planes along directions that have also a high
density of sites. In the particular case of a tilt boundary
configuration, the planes of the IL normal to the tilt boundary
constitute the model's prediction for the so called twist
boundary. Hence, in the geometrical approach, twist boundaries do
provide information about the possible directions that a
perpendicular (tilt) boundary may run along. We will hence
focus on the structure of twist planes below.

The left part of Figure \ref{low} shows the twist interface
corresponding to the example above, obtained using the approach of
Sec.\ref{david} for a cubic lattice. It is a plane normal to the
$\langle 001\rangle$ direction in the cubic $(100)/\left\langle
001\right\rangle$ twist interface. As indicated above, this plane
is a map of all the possible transverse conformations (or
trajectories) of a tilt grain boundary between two grains of
given misorientation angle. A particular trajectory is indicated
in gray. Large disks represent atomic positions, or interfacial
sites. The short and long arrows indicate the O-lattice (OL) and
coincidence sites lattice (CSL) base vectors, respectively. Note
that the interface is composed of identical domains centered at
O-points (small dots), that are points of zero strain. Atomic
domains are separated by a network of primary screw 
dislocations.\cite{1}
The intersection of these dislocations with the
boundary plane is highlighted by a square grid in the figure.

The predicted atomic positions (grayed dots) match
nearly exactly the (black) regions where $\psi$ is maximum in the
nonlinear model. The short  horizontal lines on the figure mark
the positions of dislocations. These are edge dislocations, with
Burger's vectors normal to the boundary line. A similar remarkable
agreement was obtained in every case for a variety of angles
$\theta$ chosen arbitrary, in square and hexagonal symmetries.
Non-singular interfaces are characterized by more complicated
interfacial patterns (not shown).

Since elastic strain is minimized near O-points, it is expected that
tilt interfaces {\em of low angles} $\theta$ tend to run along directions
that follow the O-lattice, preserving this way the structure of the domains
around the O-points.
The O-lattice has the same symmetry as the crystal (see Figure \ref{low}, 
left), but a distinct period $\lambda_O$, given by 
$\lambda_O=\lambda/[2\sin(\theta/2)]$ independently of the symmetry\cite{7}
($\lambda$ is the period of the crystal; $\lambda_O>\lambda$
at low angles). Therefore, the grain boundaries separating two symmetric 
grains along the vertical axis should be located at some discrete 
$x$-positions, separated from each other by $\lambda_O$ 
(Figure \ref{low}, left). On the other hand, a weakly nonlinear analysis
of Eq.(\ref{sh}) allows
the derivation of the law of motion of straight grain
boundaries.\cite{5} It is found analytically that interfaces feel
a pinning potential, periodic along the $x$-direction.
Strikingly, the nonlinear model agrees perfectly with the geometrical
construction: The periodicity of the pinning potential is
$\lambda/[2\sin(\theta/2)]$ and coincides with $\lambda_O$.
In the nonlinear model, the derivation of this formula
follows from a solubility condition that imposes particular combinations
of the modes $\vec{k}_n$. The result is formally equivalent to the
the formula (\ref{OLperio}) above.
The lines going through points of the O-lattice
thus correspond to minima of a pinning potential in the nonlinear model.

\begin{figure}[bt]
\centerline{\hspace{-.5cm}\psfig{file=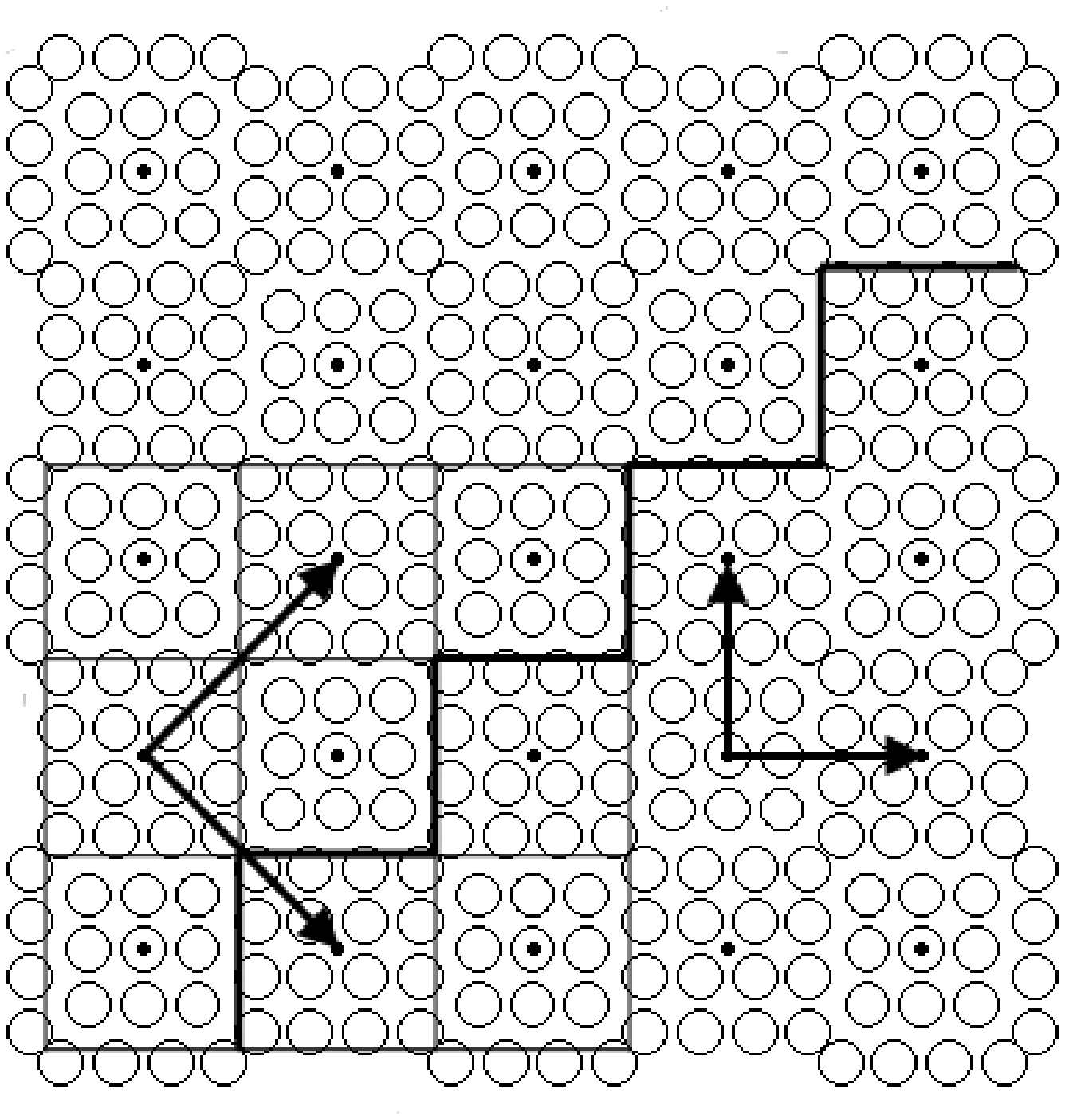,width=2.3in,angle=45}
\psfig{file=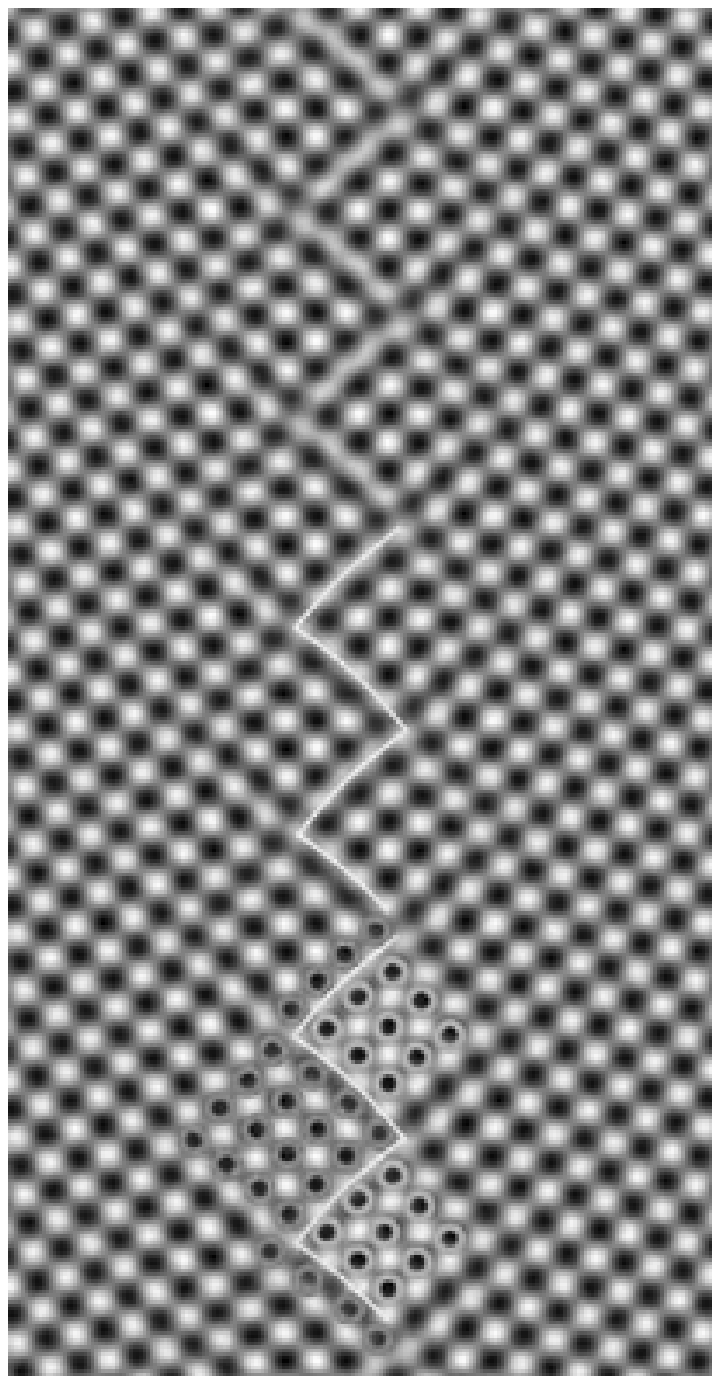,width=2.in}}
\vspace*{8pt}
\caption{\label{high} Facet formation.
Right: Nonlinear model with at $\theta=73.74^{\circ}$ ($\Sigma 25$).
Left: Geometrical construction.}
\end{figure}

Figure \ref{high} (on the right) represents a configuration obtained at time
$t=2000$ from the same nonlinear model, with a {\em high}
angle $\theta=73.74^{\circ}$.
The initially straight interface quickly destabilizes into a broken-symmetry
phase, characterized by facets of short period. This faceting instability
is not observed for low $\theta$ boundaries, for instance that of
$\theta=90-73.74$, the other symmetric boundary
orientation in this bicrystal.

The faceted structure is also visible from the geometrical
construction of $E^{\parallel}$ (left). In this configuration, a
straight tilt grain boundary would run vertically, {\it i.e.}
along the diagonal of the domains of best fit of size 3 {\em or}
4, where the distance between O-points is larger. In order to keep
the structure of these domains anyway, the interface becomes
faceted instead. It runs along the core lines of the screw
dislocation network. In this case, the interface is locally
aligned with the Burgers vectors of the dislocations that compose
it. Along this line, a repeating structure of 4 atoms facing 3
atoms of the other grain appears, as correctly reproduced by the
nonlinear method. Note that the geometrical map of Figure
\ref{high} (left) also contains the structure of symmetric tilt
boundaries with the angle $\theta=90-73.74=16.26^\circ$. 
The direction of these
interfaces is now inclined of $45^\circ$ with respect to the vertical
direction in Figure \ref{high} (left): 
the most likely trajectories are straight and run
along a line of maximum O-point density (or pinning potential minimum),
as confirmed by nonlinear calculations (not shown). If an
interface with such an orientation destabilized, it would find it
difficult to develop the same \lq\lq Chinese hat" facets as
observed for the other orientation $\theta=73.74^\circ$. 
Hence, low $\theta$ grain boundaries are probably less likely to facet. 
In conclusion, when the average orientation of a boundary is not parallel 
to the O-lattice, the boundary {\it may} become 
faceted.

Preliminary numerical results on the nonlinear model for square patterns
show that the faceting instability does not occur for
misorientation angles lower than a characteristic value
$\theta_c\approx 60^{\circ}$. This value can be interpreted
qualitatively as follows: For $\theta\ge\theta_c$, the periodicity
of the pinning potential $\lambda/[2\sin(\theta/2)]$ becomes
similar to that of the crystal making it easier for the interface
to deform under stress.

The configuration of Figure \ref{high} (right) has a relatively high elastic
stress. After sufficiently long times, it starts to slowly
evolve into more complicated patterns, with longer facet lengths
in order to relieve the initial stress.

\section{Conclusions}
\label{concl}

We have presented a theory of crystals based on a local order
parameter description and general stability and symmetry
principles. The theory is well suited for the description of short
scale phenomena, in particular the intricate spatial distributions of
atoms in grain boundaries. The results compare very well with
those given independently by a generic geometrical theory of grain
boundaries. The nonlinear nature of the model naturally
incorporate some underlying topological structures, such as the
O-lattice, which in this theory is interpreted as the set of
boundary pinning sites.

\section*{Acknowledgements}
This work was supported
by the Consejo Nacional de Ciencia y Tecnolog\'\i a (CONACYT, Mexico)
Grant number 40867-F.

\end{document}